# Bilayer Kagome Borophene with Multiple van Hove Singularities

Qian Gao[1], Qimin Yan[2], Zhenpeng Hu[1*], Lan Chen[3]

[1]School of Physics, Nankai University, Tianjin 300071, China

[2]Department of Physics, Northeastern University, Boston, MA 02115, USA

[3]Institute of Physics, Chinese Academy of Sciences, Beijing 100190, China

*Corresponding author:

Tel: +86-22-23509675 E-mail addresses: zphu@nankai.edu.cn

**Abstract:**

   The appearance of van Hove singularities near the Fermi level leads to prominent phenomena, including superconductivity, charge density wave, and ferromagnetism. Here a bilayer Kagome lattice with multiple van Hove singularities is designed and a novel borophene with such lattice (BK-borophene) is proposed by the first-principles calculations. BK-borophene, which is formed via three-center two-electron (3c–2e) σ-type bonds, is predicted to be energetically, dynamically, thermodynamically, and mechanically stable. The electronic structure hosts both conventional and high-order van Hove singularities in one band. The conventional van Hove singularity resulting from the horse saddle is 0.065 eV lower than the Fermi level, while the high-order one resulting from the monkey saddle is 0.385 eV below the Fermi level. Both the singularities lead to the divergence of electronic density of states. Besides, the high-order singularity is just intersected to a Dirac-like cone, where the Fermi velocity can reach $1.34 \times 10^6$ m/s. The interaction between the two Kagome lattices is critical for the appearance of high-order van Hove singularities. The novel bilayer Kagome borophene with rich and intriguing electronic structure offers an unprecedented platform for studying correlation phenomena in quantum material systems and beyond.

**Introduction:**

Van Hove singularities (VHSs) have important implications in condensed-matter physics, for they directly affect the electronic transitions, and play a prominent role in the electronic and thermal properties of crystalline solids, manifested as pronounced peaks in the absorption or emission spectra.[1-4] Theoretically, a Van Hove singularity near the Fermi level would increase the electronic density of states at the Fermi level, lead to the quantum Lifshitz phase transition,[5-7] and trigger quantum emergent phenomena like superconductivity, charge density wave (CDW), and ferromagnetism.[8-12] Experimentally, various structures, such as phosphorene,[12] carbon nanotube,[13] twisted bilayer graphene,[14-16], FeGe,[17] and RhSi(or CoSi),[18] etc., exhibit various anomalous phenomena related to van Hove singularities in spectroscopic measurements.

VHSs occur in various lattice structures,[19-22] of which Kagome lattice is the typical network composed of connected regular triangles sharing vertices, providing a unique platform for both Dirac cone (DC) and VHS.[23-26] The family of layered vanadium antimonides $AV_3Sb_5$ (A=K, Rb, Cs) with layered Kagome lattices has emerged as the preferable system to study electronic structure with VHS, including conventional and high-order VHSs (HOVHS).[27-30] HOVHS is the high-order critical point with power-law divergence in the band structure, and vitally promote the formation of complex quantum phases via interactions.[31-35] HOVHS has been more extensively realized by regulating the moiré-pattern graphene with several parameters like twist angle, pressure, or external fields.[36-41] VHSs have opened up possibilities for correlated phases, and enrich the quantum emergent phenomena.

Owing to the fascinating physics presented by VHS, the search for novel materials with VHSs near Fermi energy is strongly motivated. Compared to AV$_3$Sb$_5$ with layered structures, two-dimensional (2D) materials have the advantage of integration in electronic devices due to the thickness of single or several atomic layers. The change of interlayer coupling in 2D materials is an effective method for VHS implementation, like regulating stacking angle in moiré-pattern graphene.[42] However, the van der Waals stacked graphene with precise angle control is not technologically-mature yet for accurate, clean, and high-yield assembly.[43,44] A 2D material with stronger interlayer interaction and easier realization of VHS near Fermi level is attractive. Among the 2D materials, borophene is the recently emerging material with diverse phases and rich electronic properties, which has great potential for various applications.[45-49] As the lightest elemental Dirac materials, β$_{12}$ and χ$_3$ phases are the very promising 2D structures with highly metallic behavior.[50-52] By now, more 2D boron allotropes with various topologies and band structures exhibiting DCs have been proposed.[53-56] Borophene has also been proved the single-elemental superconductor with the high critical $T_c$ beyond 10 $K$ under conditions without high pressure and external strain.[57,58] Recently, the successful synthesis of bilayer borophene with strong interlayer coupling also provides the polymorphism of borophene and richness of electronic structures.[59,60] All these characters make the borophene a potential platform for intriguing electronic properties, including the implementation of VHS near the Fermi level.

Motivated by the Kagome lattice and rich polymorphism of borophene, we proposed a novel bilayer Kagome borophene with multiple VHSs near the Fermi level in this study. The bottom and top layers of atoms in the structure make up two Kagome lattices, respectively. It has an extremely stable structure and unique bonding characteristics. It owes both VHS and HOVHS in the electronic structure. Besides, it is characterized with various novel properties, such as the Dirac-like cone in electronic structure and even higher Fermi velocity than that of graphene. The rich electronic properties of bilayer Kagome borophene make it a promising candidate for various electronic applications like superconductivity, and probably trigger another interesting hotspot.

**Methods:**

In this study, first-principles calculations based on density functional theory (DFT) were performed to investigate the structural and electronic properties. The projected augmented wave (PAW) method[61] and the generalized gradient approximation (GGA) with Perdew−Burke−Ernzerhof (PBE)[62] exchange-correlation functional were employed for DFT calculations using the Vienna Ab initio Simulation Package (VASP) code[63,64]. A plane wave basis with an energy cutoff of 520 eV and 21×21×1 Monkhorst-Pack scheme[65] were set in the calculations. The energy precision was set to $10^{-5}$ eV, and the vacuum size is larger than 18 Å to avoid the interactions between two adjacent layers. The atomic positions and lattice parameters of the bilayer Kagome borophene were optimized using the conjugate gradient algorithm until the forces on each atom were less than 0.01 eV/Å. Following the structural optimization, the electronic structure was calculated. To verify the structural stability, the phonon dispersion was calculated using density functional perturbation theory (DFPT) method[66] with the Phonopy code[67] in concert with VASP. For the phonon calculation needs a much higher accuracy than normal electronic structure calculation, the converging criteria were tightened up to $10^{-8}$ eV for total energies. Biaxial stress-strain method was applied to evaluate the mechanical properties by elastic constants.[68] Ab initio molecular dynamics (AIMD) was simulated used a 5×5×1 supercell (150 atoms) with the temperature control of Nosé–Hoover thermostat,[69] and the time of the simulation was set to 5 ps with a time step of 1 fs. The projected crystal orbital Hamilton population (pCOHP) is analyzed by the program Local-Orbital Basis Suite Towards Electronic-Structure Reconstruction (LOBSTER)[70]. The solid-state adaptive natural density partitioning (SSAdNDP)[71] algorithm was used to analyze the bonding pattern of the structure, which allows the interpretation of chemical bonding in systems with transitional symmetry in terms of classical lone pairs and multi-center delocalized bonding.

**Results and discussions:**

**Bilayer Kagome lattice**

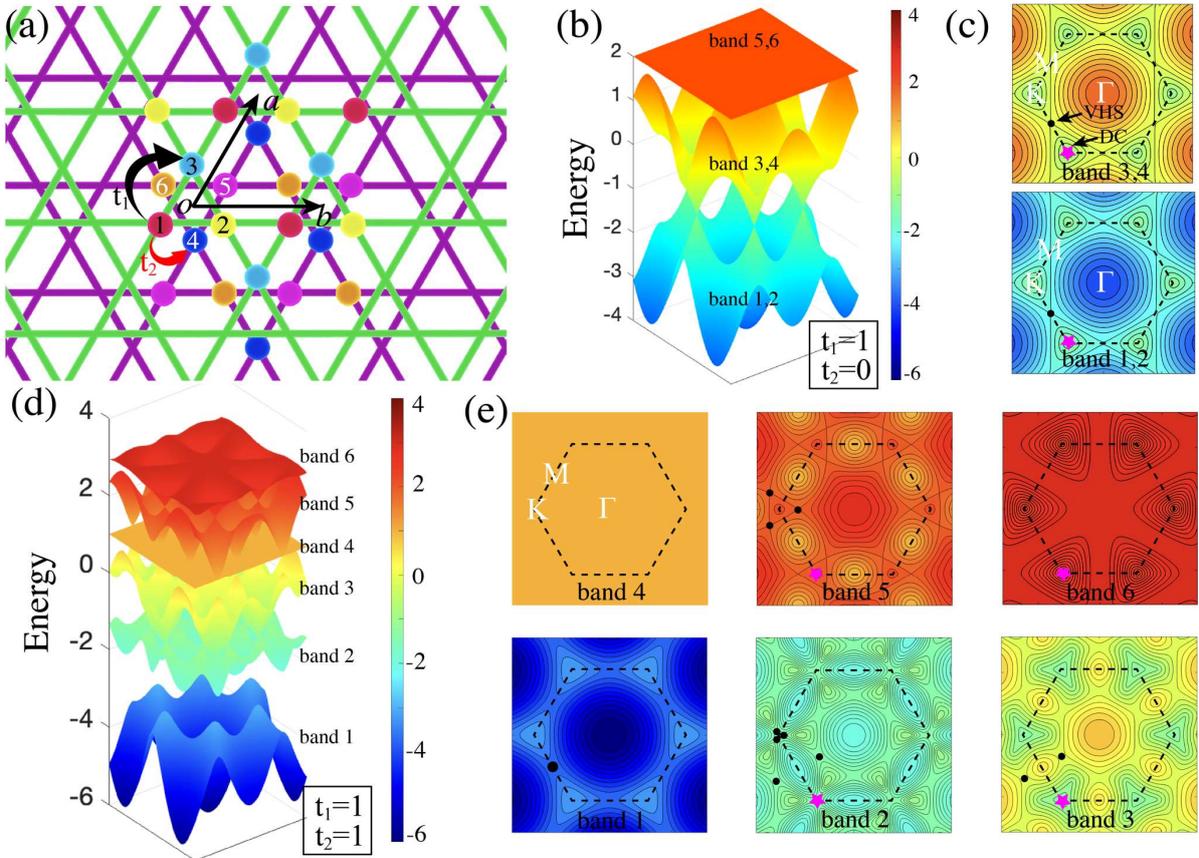

**Figure 1.** (a) Schematic diagram of bilayer Kagome lattice. The top and bottom lattices are distinguished by green and purple, respectively. Six sublattices are marked with six Arabic numerals and different colors. (b) 3D band structure and (c) energy contours of specific bands with $t_1 = 1$, and $t_2 = 0$. (d) 3D band structure and (e) energy contours of specific bands with $t_1 = 1$, and $t_2 = 1$. The same color bar is used for all 3D band structures and energy contours. The black circle and magenta pentagram represent the sites of VHS and DC, respectively.

Kagome lattice is the 2D lattice with a hexagonal unit cell consisting of three interconnected triangular sublattices. Bilayer Kagome lattice is obtained by the stacking of two Kagome lattices rotated by 60 degrees with respect to each other (Figure 1a). Unit cell of the bilayer Kagome lattice described by the basis vectors *a* and *b* comprises six sites labeled from 1 to 6. The corresponding tight-binding Hamiltonian has the following form

$$H_{tb} = -\sum_{<i,j>} t_{i,j} a_i^\dagger a_j - \sum_i \epsilon_i a_i^\dagger a_i, \tag{1}$$

where $t_{i,j}$ could be $t_1$ or $t_2$. $t_1$ and $t_2$ are the nearest-neighbor hopping amplitudes in a single Kagome lattice and between the two Kagome lattices, respectively. When only the interaction of the same layer is considered ($t_1 = 1$, $t_2 = 0$), two sets of overlapping electronic structure could be obtained (Figure 1b). The electronic structure includes DC at the *K* point, horse saddles around the *M* point as well as flat bands across the whole Brillouin zone, which are the typical Kagome band structures.[20,72] In the electronic structure, a horse saddle point corresponds to the conventional VHS.[2] Energy contours of band 1 (or 2) and band 3 (or 4) are illustrated in Figure 1c, and VHSs can be clearly distinguished from the diagram. When considering interlayer interactions ($t_1 = 1$, $t_2 \neq 0$), the originally degenerate energy bands begin to split (Figure S1). Electronic structure and energy contours with six separated energy bands ($t_1 = 1$, $t_2 = 1$) labeled from band 1 to 6 are presented in Figure 1d, e. Among them, VHS at the point *M* in band 1 always exists with the increase of $t_2$. VHS at the point *M* in band 2 and 3 gradually splits into two symmetrical VHSs with the increase of $t_2$. The *K* point in band 2 (the position of DC) generates a new singularity and gradually splits into three horse saddle points as $t_2$ increases. When three horse saddle $\propto(x^2 - y^2)$ points approach each other, they merge into a monkey saddle $\propto(x^2 - 3xy^2)$ point, which is a HOVHS (Figure S1).[26,31] Band 4 flattens as $t_2$ increases, until it eventually becomes flat band. Band 5 and Band 6 separate from the overlapping flat band and form

Dirac cone at point K, and three VHSs are also generated in Band 5. Overall, such a lattice provides an intriguing platform with multiple VHSs, and the HOVHS originates from the interaction between the two Kagome lattices.

**Bilayer Kagome Borophene**

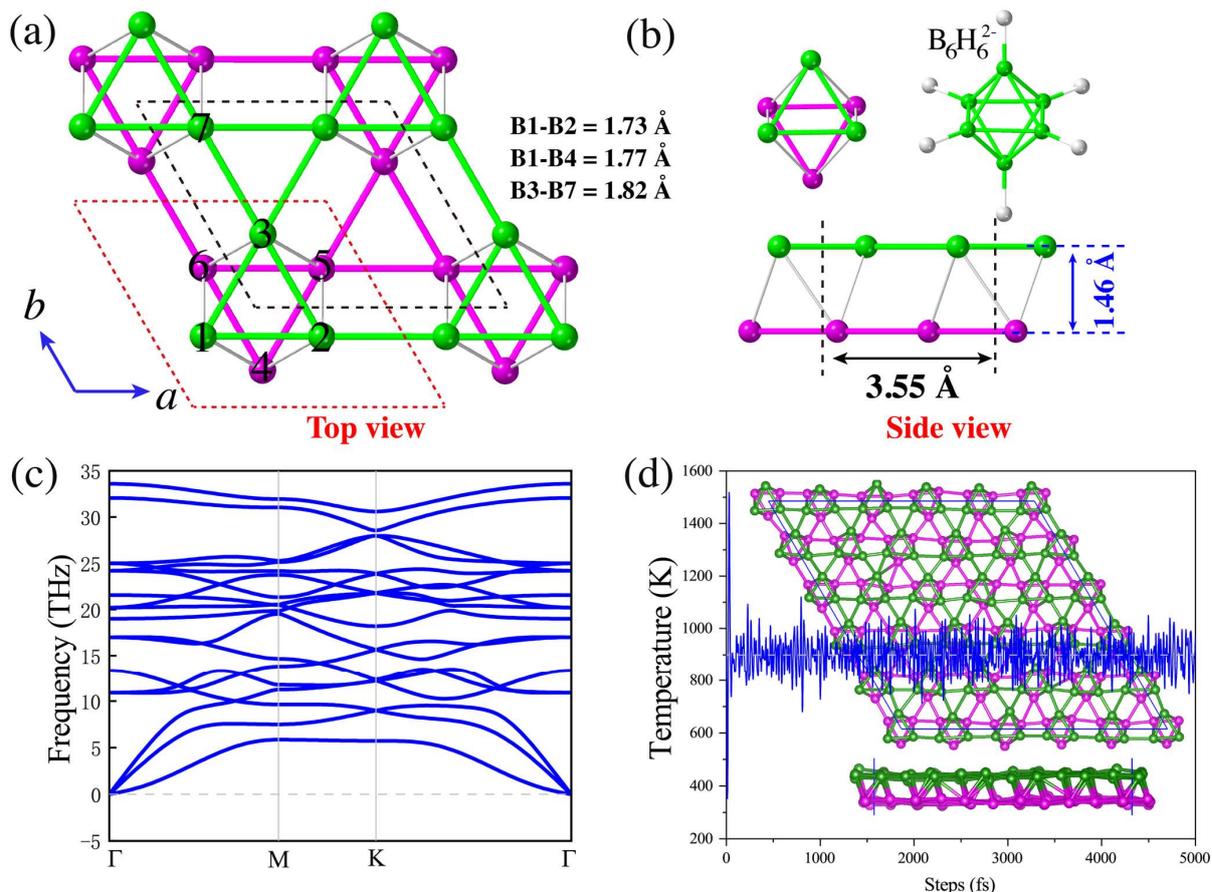

**Figure 2.** (a) Top and (b) side views of BK-borophene. It can be viewed as the stacking of two Kagome lattices rotated by 60 degrees with respect to each other, or as a configuration based on octahedral units. The dashed red lines represent the structural unit. (c) Phonon band structures along high symmetric points in Brillouin zone. (d) Molecular dynamics simulation at $T = 900$ K with 5000 fs. A color coding is used to distinguish the top and bottom atoms.

Bilayer Kagome borophene (BK-borophene) is proposed by replacing the sublattices with boron atoms. After structural optimization, the stable structure of a novel boron allotrope is finally obtained (Figure 2a and 2b). The structure has a lattice constant of 3.55 Å, and is characterized by three bonding types, which have the length of 1.82 (bond 1), 1.73 (bond 2), and 1.77 (bond 3) Å. All atoms are six coordinated, forming two absolute planes. The distance between the two planes is 1.46 Å. From another point of view, the structure can also be viewed as combination of octahedron structures, whose unit cell is illustrated by the dashed red lines in Figure 2a and left top view in Figure 2b. In fact, as the smallest known hydro-closo-borate, octahedral $[B_6H_6]^{2-}$ (right top view in Figure 2b) has been prepared long ago and comprehensive investigated,[73] indicating the possible synthesis of BK-borophene through polymerization reactions.

The energy of BK-borophene is -6.14 eV per atom, which is much close to that of $β_{12}$ (-6.25 eV/atom) and $χ_3$ phases (-6.28 eV/atom) in our calculations. The four 2D elastic constants are $C_{11} = C_{22} = 251.69$, $C_{12} = 32.18$, and $C_{66} = 109.76 \text{ N/m}$, which satisfy the Born-Huang criteria of the mechanical stability.[74] The 2D Young's modulus ($Y$) in the Cartesian [10] and [01] directions are

$$Y_{[10]} = \frac{C_{11}C_{22} - C_{12}^2}{C_{22}}, \text{ and } Y_{[01]} = \frac{C_{11}C_{22} - C_{12}^2}{C_{11}}, \qquad (2)$$

the shear modulus ($G$) is

$$G = C_{66}, \qquad (3)$$

and Poisson's ratios ($v$) are given as[75]:

$$v_{[10]} = \frac{C_{12}}{C_{22}}, \text{ and } v_{[01]} = \frac{C_{12}}{C_{11}}. \qquad (4)$$

According to the formulas, the Young's and shear modulus of BK-borophene are smaller than that of graphene but larger than $β_{12}$ or $χ_3$ phases (Table 1). The phonon spectrum of the structure is presented in Figure 2c, showing no imaginary frequencies, indicating its dynamical stability. The highest frequency observed in the phonon spectrum is about 34 THz, which could be attributed to the high stiffness of boron bonds. AIMD simulations were performed to check the thermodynamical stability, as shown in Figure 2d and S2. BK-borophene is stable under the temperature below 900 K. When the temperature reached 1000 K, the system experienced bond fracture and reconstruction (Figure S2). AIMD simulations reveal that the structure remains thermodynamically stable even in the environments with high-temperature.

It is noticed that a two-dimensional metallic boron (2D-$B_6$) constructed by $B_6$ octahedron was already proposed in previous study,[76] which is a quasi-2D structure formed by connecting the four vertices of octahedron in a square lattice (Figure S3). The energy of BK-borophene is lower by 0.40 eV/atom than 2D-$B_6$, indicating that BK-borophene could be easier to synthesize in experiments. For comparison, the data of 2D-$B_6$ including Young's modulus, Shear modulus, as well as Poisson's ratio are also listed in Table 1. Overall, BK-borophene has great energetic, mechanical, kinetic and thermodynamic stability.

Table 1. Energies per atom ($E$), and mechanical properties of Young's modulus in the Cartesian [10] and [01] directions ($Y_{[10]}/Y_{[01]}$), 2D shear modulus ($G$), and Poisson's ratio in the Cartesian [10] and [01] directions ($v_{[10]}/v_{[01]}$) for BK-borophene, graphene, $\beta_{12}$, $\chi_3$, and 2D-$B_6$. Asterisk (*) represents the results from literature.[76]

| Materials | $E$ (eV/atom) | $Y_{[10]}/Y_{[01]}$ (N/m) | $G$ (N/m) | $v_{[10]}/v_{[01]}$ |
|---|---|---|---|---|
| BK-borophene | -6.14 | 247.57/247.57 | 109.76 | 0.128/0.128 |
| Graphene | / | 344.83/344.83 | 147.74 | 0.175/0.175 |
| $\beta_{12}$ | -6.25 | 186.80/217.92 | 68.44 | 0.152/0.178 |
| $\chi_3$ | -6.28 | 203.00/196.23 | 57.01 | 0.136/0.131 |
| 2D-$B_6$ | -5.74 | 154.02/154.02 | 5.99 | -0.087/-0.087 |
| | -5.76* | 149.01* | | -0.08* |

**Electronic structure**

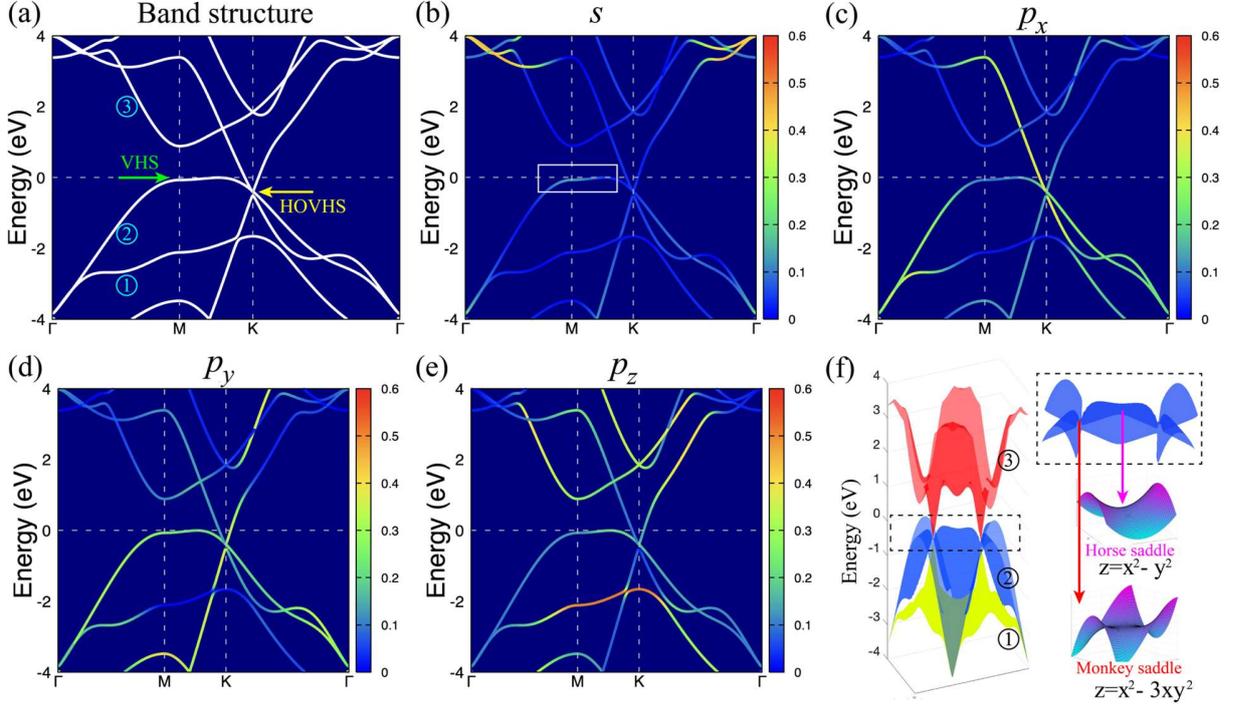

**Figure 3**. (a) Electronic band structure along symmetry lines of the Brillouin zone based on the PBE level, and (b-e) $s$, $p$ projected band structure. (f) 3D plot of the bands 1-3 with three colors (green, blue, and red), and comparison of horse saddle and monkey saddle according to the mathematical functions $z = x^2 - y^2$ (horse saddle), and $z = x^2 - 3xy^2$ (monkey saddle). The Fermi level is set at 0 eV.

The $\beta_{12}$ or $\chi_3$ phases of borophene that have been synthesized exhibit metallic properties,[45] and BK-borophene is no exception. Figure 3 illustrates the band structure of BK-borophene. The bands near Fermi level from low to high energy are marked as bands 1-3 (Figure 3a). In band 2, two connected horse saddle and monkey saddle appear, and the connection is just in contact with the Fermi level. The horse saddle point (VHS) and monkey saddle point (HOVHS) are located at the energies of 0.065 eV (*M*) and 0.385 eV (*K*) below the Fermi level, respectively. According to various studies, once a VHS approaches Fermi energy, various correlated electronic phases, like ferromagnetism,[77] CDW,[78] or superconductivity[79] would be substantially enhanced. At the monkey saddle point, a Dirac-like cone also appears. Figures 3b-e show the contributions of different orbitals (including $s$, $p_x$, $p_y$, and $p_z$) of the boron atoms to the band structure. The $p_y$ orbital has significant contribution to the bands near the Fermi level, while the $s$ orbital of boron atoms has almost no contribution to the band near the Fermi level, except for a slight contribution around the horse saddle. Figure 3f presents the 3D structure of band 1-3, so that the horse saddle and monkey saddle points can be clearly distinguished. Mathematically, the horse and monkey saddles are surfaces defined by the Cartesian equations $z \propto x^2 - y^2$, and $z \propto x^2 - 3xy^2$, consistent with the structures appearing in the 3D energy band.

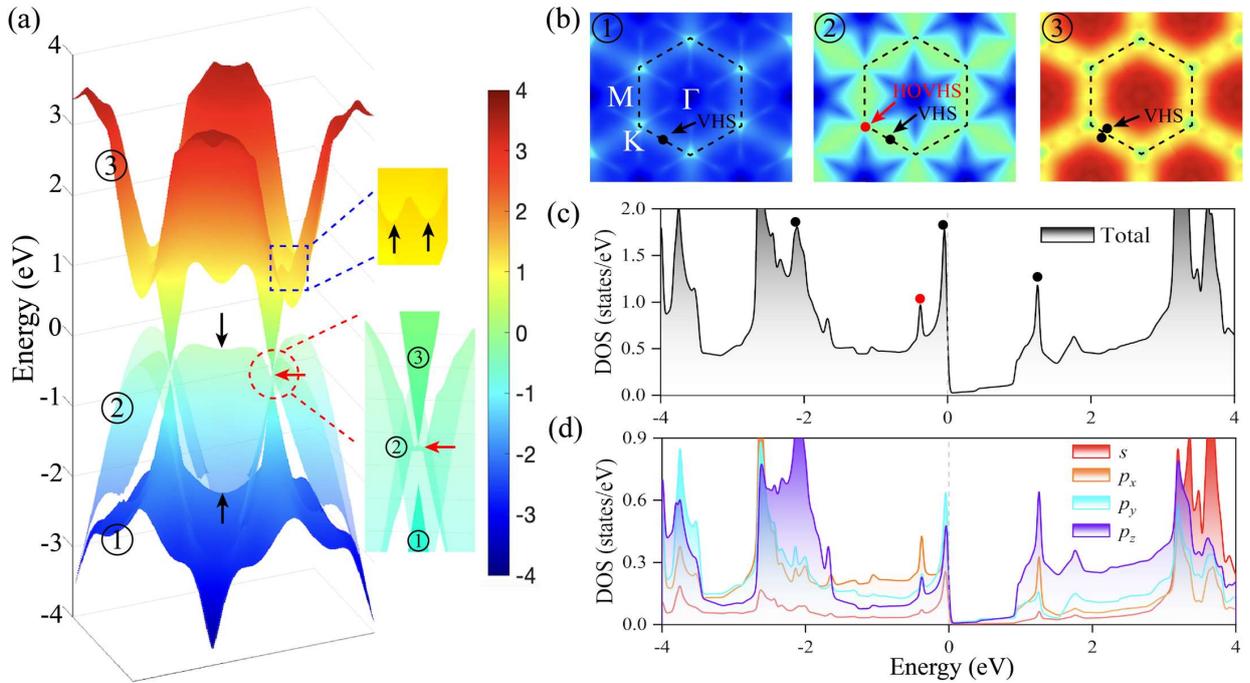

**Figure 4.** (a) 3D structure of band 1-3. The black and red arrows separately point to VHS and HOVHS, respectively. (b) The energy contours of bands 1-3. (c) Total and (d) projected density of states (PDOS) based on the PBE level. The black and red circles represent the sites of VHS and HOVHS, respectively. The dashed black lines show the Brillouin zone. The same color bar is used for 3D band structures and energy contours. The Fermi level is set at 0 eV.

To better analyze the singularities, we have carried out a detailed analysis on the 3D band structure, as shown in Figure 4. In addition to the band mentioned earlier that appears in band 2, VHSs also appears at the energy of 1.202 eV higher than Fermi level in band 3 (in the blue dashed box in Figure 4a), which is not in the highly symmetric path of the Brillouin zone. The contour plots of band 1-3 in Figure 4b better illustrated the sites of VHSs in Brillouin zone. Interestingly, as presented in the zoomed circle in Figure 4a, it could be found that the two Dirac-like bands 1 and 3 are not touched, but rather the bottom of band 3 is precisely connected to the saddle point of band 2, exactly consistent with the situation of band 2 and 3 in the tight-binding model ($t_1 = 1$, $t_2 \neq 0$). The energy gap between band 1 and 3 is 0.117 eV. The Fermi velocity in band 3 at the K point can reach $1.34 \times 10^6$ m/s, even higher than that of graphene ($1.10 \times 10^6$ m/s)[80]. When considering spin-orbit coupling, a band gap of 0.017 eV was opened between bands 2 and 3, and the energy gap between bands 1 and 3 is enlarged to 0.138 eV.

VHS would lead to a divergence in the density of states (DOS), showing the pronounced peaks in Figure 4c. These peaks signify the existence of localized electronic states and are indicative of the unique topological nature of the BK-borophene. The peaks resulting from VHS and HOVHS near the Fermi level are obviously presented. HOVHS at monkey saddle is a multi-critical Lifshitz point, which portends a tendency towards many-body instabilities.[31,41] The VHS-induced Lifshitz transitions could be observed in spectral probes of the electronic bands, such as angle-resolved photoemission spectroscopy or Landau level spectroscopy.[31] To gain insight into the orbital contributions, we performed orbital projection analysis of density of states (PDOS), as illustrated in Figure 4d. The results of PDOS indicate the entire electronic structure is contributed by both the *s* and *p* orbitals, indicating the hybridization of *s* and *p* orbitals. Near the Fermi level, the DOS is predominantly influenced by the *p* orbitals. For VHS in band 3, the $p_z$ orbital contributes the most to the DOS. This observation suggests the presence of complex electronic interactions and highlights the intricate nature of the bonding in BK-borophene.

**Bonding analysis**

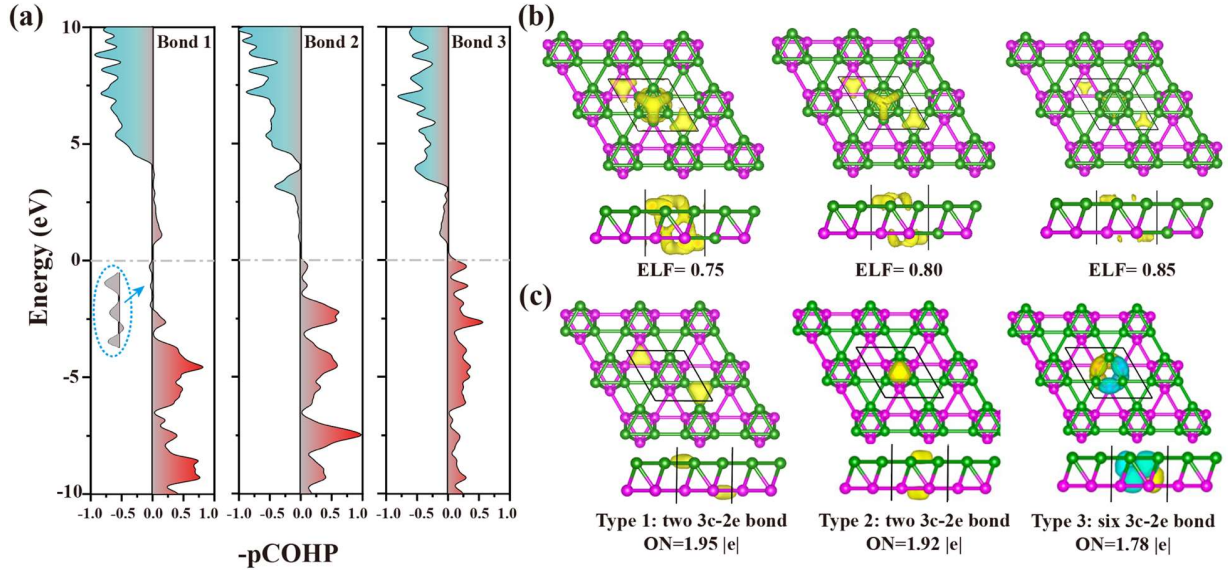

**Figure 5.** (a) The -pCOHP plots of the three B-B bonds, (b) isosurfaces of the bilayer Kagome borophene with ELF=0.75, 0,80, and 0.85, respectively. (c) Bonding orbitals based on SSAdNDP method, including the occupations of electrons and types of bonding.

Figure 5a displays the information of bonding and antibonding states. Below the Fermi level, no antibonding states are present in bond 2 and bond 3, whereas bond 1 shows little antibonding states below the Fermi level. As a convenient descriptor of electron fluctuation, the electron localization function (ELF) could characterize the chemical bond (Figure 5b). From the isosurface plots at ELF=0.75, electron localization is observed in all the triangular regions formed by boron atoms; at ELF=0.8, the localized electrons between the top and bottom boron layers disappear; at ELF=0.85, only the triangular region formed by bond 1 exhibits noticeable electron localization. The localized nature of the electrons at the triangular sites of bilayer Kagome borophene indicates a character of three center bonds. To confirm this, the bonding pattern was also examined, as depicted in Figure 5c. The structure consists of ten 3c-2e bonds with three types: two with electron occupancy of 1.95 $|e|$; two with electron occupancy of 1.92 $|e|$; six with electron occupancy of 1.78

|*e*|. Unlike traditional covalent bonds where electrons are localized between two atoms, 3c-2e bonds involve the delocalization of electrons over all three atoms involved. The delocalization allows for the sharing of electron density across the bond, resulting in stabilization of the system.

**Conclusion:**

We have designed a bilayer Kagome lattice and proposed a novel bilayer Kagome borophene with multiple van Hove singularities by the first-principles calculations. The stability of such structure is confirmed through total energy per atom, phonon dispersion, molecular dynamic simulations, as well as elastic constants calculations. There are ten three-center two-electron (3c–2e) σ-type bonds in the unit cell, which allows for the sharing of electron density across the bond, resulting in stabilization of the structure. The electronic structure hosts both conventional and high-order VHS. VHS resulting from the horse saddle is 0.065 eV lower than Fermi level, while HOVHS resulting from the monkey saddle appear about 0.385 eV below the Fermi level. Both VHS and HOVHS lead to the divergence of electronic density of states. Besides, HOVHS is just intersected to a Dirac-like cone, and the spin-orbit coupling effect opens a band gap of about 17 meV. The Fermi velocity can reach $1.34 \times 10^6$ m/s, even higher than that of graphene ($1.10 \times 10^6$ m/s). All the results showing KB-borophene is a rare example of bilayer Kagome lattice which host VHS, HOVHS, as well as Dirac-like cone around the Fermi level, presenting an opportunity to establish the connection between various phenomena. These features make BK-borophene a promising candidate for applications in electronics, as well as other fields. We hope that this study will stimulate further theoretical and experimental research of correlation phenomena in this system.

**Acknowledgements**

This work is partially supported by the National Natural Science Foundation of China (12134019，21773124), by the Fundamental Research Funds for the Central Universities (Nankai University, Grant No. 63213042, 63221346, and ZB22000103). This work is also supported by Supercomputing Center of Nankai University (NKSC) and TianHe-2A at the National Supercomputer Center, Guangzhou.

**Conflict of Interest**

The authors declare that they have no conflict of interest.

**Data Availability Statement**

The data that support the findings of this study are available from the corresponding author upon reasonable request.

**References:**

1   Zhang, J. *et al.* Observation of Van Hove Singularities and Temperature Dependence of Electrical Characteristics in Suspended Carbon Nanotube Schottky Barrier Transistors. *Nano-Micro Lett.* **10**, 25 (2017).
2   Van Hove, L. The Occurrence of Singularities in the Elastic Frequency Distribution of a Crystal. *Phys. Rev.* **89**, 1189-1193 (1953).
3   Ma, C. *et al.* Evidence of van Hove Singularities in Ordered Grain Boundaries of Graphene. *Phys. Rev. Lett.* **112**, 226802 (2014).
4   Houssa, M. & Ausloos, M. Thermal conductivity of high-Tc superconductors: effect of Van Hove singularities. *Physica C* **265**, 258-266 (1996).
5   Hornreich, R. M. The Lifshitz point: Phase diagrams and critical behavior. *J. Magn. Magn. Mater.* **15-18**, 387-392 (1980).
6   Volovik, G. E. Topological Lifshitz transitions. *Low Temp. Phys.* **43**, 47-55 (2017).
7   Mohan, P. & Rao, S. Interplay of Floquet Lifshitz transitions and topological transitions in bilayer Dirac materials. *Phys. Rev. B* **98**, 165406 (2018).
8   Uldemolins, M., Mesaros, A. & Simon, P. Effect of Van Hove singularities on Shiba states in two-dimensional *s*-wave superconductors. *Phys. Rev. B* **103**, 214514 (2021).
9   Vozmediano, M. A. H., González, J., Guinea, F., Alvarez, J. V. & Valenzuela, B. Properties of electrons near a Van Hove singularity. *J. Phys. Chem. Solids* **63**, 2295-2297 (2002).
10  Chen, K. S. *et al.* Role of the van Hove singularity in the quantum criticality of the Hubbard model. *Phys. Rev. B* **84**, 245107 (2011).
11  Wu, Z., Wu, Y.-M. & Wu, F. Pair density wave and loop current promoted by Van Hove singularities in moiré systems. *Phys. Rev. B* **107**, 045122 (2023).

12  Ziletti, A., Huang, S. M., Coker, D. F. & Lin, H. Van Hove singularity and ferromagnetic instability in phosphorene. *Phys. Rev. B* **92**, 085423 (2015).
13  Sfeir, M. Y. *et al.* Probing Electronic Transitions in Individual Carbon Nanotubes by Rayleigh Scattering. *Science* **306**, 1540-1543 (2004).
14  Li, G. *et al.* Observation of Van Hove singularities in twisted graphene layers. *Nat. Phys.* **6**, 109-113 (2010).
15  Isobe, H., Yuan, N. F. Q. & Fu, L. Unconventional Superconductivity and Density Waves in Twisted Bilayer Graphene. *Phys. Rev. X* **8**, 041041 (2018).
16  Wu, S., Zhang, Z., Watanabe, K., Taniguchi, T. & Andrei, E. Y. Chern insulators, van Hove singularities and topological flat bands in magic-angle twisted bilayer graphene. *Nat. Mater.* **20**, 488-494 (2021).
17  Teng, X. *et al.* Magnetism and charge density wave order in kagome FeGe. *Nat. Phys.* **19**, 814–822 (2023).
18  Sanchez, D. S. *et al.* Tunable topologically driven Fermi arc van Hove singularities. *Nat. Phys.* **19**, 682–688 (2023).
19  Kulynych, Y. & Oriekhov, D. O. Differential entropy per particle as a probe of van Hove singularities and flat bands. *Phys. Rev. B* **106**, 045115 (2022).
20  Lin, Y.-P. & Nandkishore, R. M. Complex charge density waves at Van Hove singularity on hexagonal lattices: Haldane-model phase diagram and potential realization in the kagome metals $AV_3Sb_5$ (A=K, Rb, Cs). *Phys. Rev. B* **104**, 045122 (2021).
21  Yudin, D. *et al.* Fermi Condensation Near van Hove Singularities Within the Hubbard Model on the Triangular Lattice. *Phys. Rev. Lett.* **112**, 070403 (2014).
22  Zhang, S. *et al.* Kagome bands disguised in a coloring-triangle lattice. *Phys. Rev. B* **99**, 100404 (2019).
23  Kiesel, M. L., Platt, C. & Thomale, R. Unconventional Fermi Surface Instabilities in the Kagome Hubbard Model. *Phys. Rev. Lett.* **110**, 126405 (2013).
24  Park, T., Ye, M. & Balents, L. Electronic instabilities of kagome metals: Saddle points and Landau theory. *Phys. Rev. B* **104**, 035142 (2021).
25  Yin, J.-X., Lian, B. & Hasan, M. Z. Topological kagome magnets and superconductors. *Nature* **612**, 647-657 (2022).
26  Scammell, H. D., Ingham, J., Li, T. & Sushkov, O. P. Chiral excitonic order from twofold van Hove singularities in kagome metals. *Nat. Commun.* **14**, 605 (2023).
27  Hu, Y. *et al.* Rich nature of Van Hove singularities in Kagome superconductor $CsV_3Sb_5$. *Nat. Commun.* **13**, 2220 (2022).
28  Kang, M. *et al.* Twofold van Hove singularity and origin of charge order in topological kagome superconductor $CsV_3Sb_5$. *Nat. Phys.* **18**, 301-308 (2022).
29  Neupert, T., Denner, M. M., Yin, J.-X., Thomale, R. & Hasan, M. Z. Charge order and superconductivity in kagome materials. *Nat. Phys.* **18**, 137-143 (2022).
30  Wu, X. *et al.* Nature of Unconventional Pairing in the Kagome Superconductors $AV_3Sb_5$ (A=K, Rb, Cs). *Phys. Rev. Lett.* **127**, 177001 (2021).
31  Shtyk, A., Goldstein, G. & Chamon, C. Electrons at the monkey saddle: A multicritical Lifshitz point. *Phys. Rev. B* **95**, 035137 (2017).
32  Classen, L., Chubukov, A. V., Honerkamp, C. & Scherer, M. M. Competing orders at higher-order Van Hove points. *Phys. Rev. B* **102**, 125141 (2020).
33  Efremov, D. V. *et al.* Multicritical Fermi Surface Topological Transitions. *Phys. Rev. Lett.* **123**, 207202 (2019).

34  Lin, Y.-P. & Nandkishore, R. M. Parquet renormalization group analysis of weak-coupling instabilities with multiple high-order Van Hove points inside the Brillouin zone. *Phys. Rev. B* **102**, 245122 (2020).

35  Aksoy, Ö. M. *et al.* Single monkey-saddle singularity of a Fermi surface and its instabilities. *Phys. Rev. B* **107**, 205129 (2023).

36  Yuan, N. F. Q. & Fu, L. Classification of critical points in energy bands based on topology, scaling, and symmetry. *Phys. Rev. B* **101**, 125120 (2020).

37  Ichinokura, S. *et al.* Van Hove singularity and Lifshitz transition in thickness-controlled Li-intercalated graphene. *Phys. Rev. B* **105**, 235307 (2022).

38  Xu, S. *et al.* Tunable van Hove singularities and correlated states in twisted monolayer–bilayer graphene. *Nat. Phys.* **17**, 619-626 (2021).

39  Turkel, S. *et al.* Orderly disorder in magic-angle twisted trilayer graphene. *Science* **376**, 193-199 (2022).

40  Brihuega, I. *et al.* Unraveling the Intrinsic and Robust Nature of van Hove Singularities in Twisted Bilayer Graphene by Scanning Tunneling Microscopy and Theoretical Analysis. *Phys. Rev. Lett.* **109**, 196802 (2012).

41  Yuan, N. F. Q., Isobe, H. & Fu, L. Magic of high-order van Hove singularity. *Nat. Commun.* **10**, 5769 (2019).

42  Seiler, A. M. *et al.* Quantum cascade of correlated phases in trigonally warped bilayer graphene. *Nature* **608**, 298-302 (2022).

43  Chu, Y. *et al.* A review of experimental advances in twisted graphene moiré superlattice*. *Chin. Phys. B* **29**, 128104 (2020).

44  Nimbalkar, A. & Kim, H. Opportunities and Challenges in Twisted Bilayer Graphene: A Review. *Nano-Micro Lett.* **12**, 126 (2020).

45  Zhang, Z., Penev, E. S. & Yakobson, B. I. Two-dimensional boron: structures, properties and applications. *Chem. Soc. Rev.* **46**, 6746-6763 (2017).

46  Li, Q. *et al.* Synthesis of borophane polymorphs through hydrogenation of borophene. *Science* **371**, 1143-1148 (2021).

47  Kaneti, Y. V. *et al.* Borophene: Two-dimensional Boron Monolayer: Synthesis, Properties, and Potential Applications. *Chem. Rev.* **122**, 1000-1051 (2022).

48  Feng, B. *et al.* Experimental realization of two-dimensional boron sheets. *Nat. Chem.* **8**, 563-568 (2016).

49  Ou, M. *et al.* The Emergence and Evolution of Borophene. *Adv. Sci.* **8**, 2001801 (2021).

50  Ezawa, M. Triplet fermions and Dirac fermions in borophene. *Phys. Rev. B* **96**, 035425 (2017).

51  Feng, B. *et al.* Dirac Fermions in Borophene. *Phys. Rev. Lett.* **118**, 096401 (2017).

52  Feng, B. *et al.* Discovery of 2D Anisotropic Dirac Cones. *Adv. Mater.* **30**, 1704025 (2018).

53  Yi, W. C. *et al.* Honeycomb Boron Allotropes with Dirac Cones: A True Analogue to Graphene. *J. Phys. Chem. Lett.* **8**, 2647-2653 (2017).

54  Ma, F. *et al.* Graphene-like Two-Dimensional Ionic Boron with Double Dirac Cones at Ambient Condition. *Nano Lett.* **16**, 3022-3028 (2016).

55  Kou, L. *et al.* Auxetic and Ferroelastic Borophane: A Novel 2D Material with Negative Possion's Ratio and Switchable Dirac Transport Channels. *Nano Lett.* **16**, 7910-7914 (2016).

56  Wu, R. *et al.* Micrometre-scale single-crystalline borophene on a square-lattice Cu(100)

57  Zhao, Y., Zeng, S. & Ni, J. Superconductivity in two-dimensional boron allotropes. *Phys. Rev. B* **93**, 014502 (2016).
58  Penev, E. S., Kutana, A. & Yakobson, B. I. Can Two-Dimensional Boron Superconduct? *Nano Lett.* **16**, 2522-2526 (2016).
59  Chen, C. *et al.* Synthesis of bilayer borophene. *Nat. Chem.* **14**, 25-31 (2022).
60  Liu, X. *et al.* Borophene synthesis beyond the single-atomic-layer limit. *Nat. Mater.* **21**, 35-40 (2022).
61  Blöchl, P. E. Projector augmented-wave method. *Phys. Rev. B* **50**, 17953-17979 (1994).
62  Perdew, J. P., Burke, K. & Ernzerhof, M. Generalized Gradient Approximation Made Simple. *Phys. Rev. Lett.* **77**, 3865-3868 (1996).
63  Kresse, G. & Furthmüller, J. Efficiency of ab-initio total energy calculations for metals and semiconductors using a plane-wave basis set. *Comput. Mater. Sci.* **6**, 15-50 (1996).
64  Kresse, G. & Furthmüller, J. Efficient iterative schemes for ab initio total-energy calculations using a plane-wave basis set. *Phys. Rev. B* **54**, 11169-11186 (1996).
65  Monkhorst, H. J. & Pack, J. D. Special points for Brillouin-zone integrations. *Phys. Rev. B* **13**, 5188-5192 (1976).
66  Baroni, S., de Gironcoli, S., Dal Corso, A. & Giannozzi, P. Phonons and related crystal properties from density-functional perturbation theory. *Rev. Mod. Phys.* **73**, 515-562 (2001).
67  Togo, A. & Tanaka, I. First principles phonon calculations in materials science. *Scr. Mater.* **108**, 1-5 (2015).
68  Roundy, D. & Cohen, M. L. Ideal strength of diamond, Si, and Ge. *Phys. Rev. B* **64**, 212103 (2001).
69  Martyna, G. J., Klein, M. L. & Tuckerman, M. Nosé–Hoover chains: The canonical ensemble via continuous dynamics. *J. Chem. Phys.* **97**, 2635-2643 (1992).
70  Maintz, S., Deringer, V. L., Tchougréeff, A. L. & Dronskowski, R. LOBSTER: A tool to extract chemical bonding from plane-wave based DFT. *J. Comput. Chem.* **37**, 1030-1035 (2016).
71  Galeev, T. R., Dunnington, B. D., Schmidt, J. R. & Boldyrev, A. I. Solid state adaptive natural density partitioning: a tool for deciphering multi-center bonding in periodic systems. *Phys. Chem. Chem. Phys.* **15**, 5022-5029 (2013).
72  Kiesel, M. L. & Thomale, R. Sublattice interference in the kagome Hubbard model. *Phys. Rev. B* **86**, 121105 (2012).
73  Preetz, W. & Peters, G. The Hexahydro-closo-hexaborate Dianion $[B_6H_6]^{2-}$ and Its Derivatives. *Eur. J. Inorg. Chem.* **1999**, 1831-1846 (1999).
74  Ding, Y. & Wang, Y. Density Functional Theory Study of the Silicene-like SiX and $XSi_3$ (X = B, C, N, Al, P) Honeycomb Lattices: The Various Buckled Structures and Versatile Electronic Properties. *J. Phys. Chem. C* **117**, 18266-18278 (2013).
75  Andrew, R. C., Mapasha, R. E., Ukpong, A. M. & Chetty, N. Mechanical properties of graphene and boronitrene. *Phys. Rev. B* **85** (2012).
76  Tkachenko, N. V. *et al.* Superoctahedral two-dimensional metallic boron with peculiar magnetic properties. *Phys. Chem. Chem. Phys.* **21**, 19764-19771 (2019).
77  Hlubina, R., Sorella, S. & Guinea, F. Ferromagnetism in the Two Dimensional *t-t'* Hubbard Model at the Van Hove Density. *Phys. Rev. Lett.* **78**, 1343-1346 (1997).
78  Nayak, C. Density-wave states of nonzero angular momentum. *Phys. Rev. B* **62**, 4880-

|     | 4889 (2000). |
| --- | --- |
| 79  | Honerkamp, C. & Salmhofer, M. Magnetic and Superconducting Instabilities of the Hubbard Model at the Van Hove Filling. *Phys. Rev. Lett.* **87**, 187004 (2001). |
| 80  | Zhang, Y., Tan, Y.-W., Stormer, H. L. & Kim, P. Experimental observation of the quantum Hall effect and Berry's phase in graphene. *Nature* **438**, 201-204 (2005). |